\documentclass[twocolumn,showpacs,prl,amsmath,amssymb]{revtex4}

\usepackage{graphicx}
\usepackage{dcolumn}
\usepackage{bm}
\usepackage{epsfig} 

\newcommand{\eps}  {\varepsilon}

\let\vev\VEV
\def\baselinestretch{1.2}
\def\roughly#1{\mathrel{\raise.3ex\hbox{$#1$\kern-.75em
      \lower1ex\hbox{$\sim$}}}} 

\def\e6{E(6)}
\def\10{SO(10)}

\def\3lrBL{$SU(3)_{c}\otimes SU(2)_L \otimes SU(2)_R \otimes U(1)_{B-L}$}
\def\422{SU(4) $\otimes$ SU(2) $\otimes$ SU(2)}
\newcommand{\AddrLisb}{%
 Departamento de F\'\i sica and CFIF, Instituto Superior T\'ecnico\\
          Av. Rovisco Pais 1, $\:$ 1049-001 Lisboa, Portugal }
\newcommand{\AHEP}{Instituto de F\'{\i}sica Corpuscular --
  C.S.I.C./Universitat de Val{\`e}ncia \\
  Campus de Paterna, Apt 22085,
  E--46071 Val{\`e}ncia, Spain}
\newcommand{\guj}{%
Physical Research Laboratory, Ahmedabad 380 009, India}

\begin{document}

\preprint{IFIC/06-25}

\title{Thermal leptogenesis in extended supersymmetric seesaw}

\author{ M. Hirsch}
\affiliation{\AHEP}
\author{M. Malinsk\'y}
\email{malinsky@phys.soton.ac.uk}
\affiliation{School of Physics and Astronomy, University of Southampton, 
SO16 1BJ Southampton, United Kingdom}
\author{J. C. Rom\~ao}\email{jorge.romao@ist.utl.pt} 
\affiliation{\AddrLisb}
\author{U. Sarkar}
\affiliation{\guj}
\author{J. W. F. Valle}
\email{valle@ific.uv.es}
\homepage{http://ahep.uv.es}
\affiliation{\AHEP}

\date{\today}

\begin{abstract} 
  We consider an extended supersymmetric SO(10) seesaw model with only
  doublet Higgs scalars, in which neutrino masses are suppressed by
  the scale of D-parity violation. Leptogenesis can occur at the TeV
  scale through the decay of a singlet $\Sigma$, thereby avoiding the
  gravitino crisis.  Washout of the asymmetry can be effectively
  suppressed by the absence of direct couplings of $\Sigma$ to
  leptons.
\end{abstract}

\maketitle



One of the most attractive scenarios to account for the
baryon-to-photon ratio of the universe is leptogenesis
\cite{Fukugita:1986hr,Buchmuller:2005eh} in the context of the
seesaw-mechanism~\cite{Minkowski:1977sc,Orloff:2005nu,schechter:1980gr,schechter:1982cv,Lazarides:1980nt}.
According to this the out-of-equilibrium decays of the lightest
right-handed (s)neutrinos produce a net lepton number which is later
reprocessed into the observed baryon asymmetry.
However, if they are thermally produced in the early universe, the
reheating temperature ($T_R$) should exceed $\sim 10^{9}$ GeV
\cite{Buchmuller:2004nz,Giudice:2003jh}. In the context of
supersymmetry, this leads to the overproduction of gravitinos, with
catastrophic consequences for the evolution of the
universe~\cite{Khlopov:1984pf}.  Although somewhat model-dependent,
the upper bound on $T_R$ from gravitino overproduction is rather
stringent and can be as strong as $T_R< 10^{6-7}$
GeV~\cite{Kawasaki:2004qu}.

Here we suggest a way to cure this problem by extending the seesaw
mechanism. In contrast to Ref.~\cite{Farzan:2005ez} we keep R-parity
conserved and adopt a grand unified SO(10) model, already proposed in
\cite{Malinsky:2005bi}. The model requires three sequential gauge
singlet superfields $S_i, i = 1,2,3$ in addition to the three usual
fermions in the 16-dimensional representation of
SO(10)~\cite{mohapatra:1986bd}.  In contrast to conventional seesaw,
the left-right symmetry is broken only by Higgs
doublets~\cite{Akhmedov:1995vm,Barr:2005ss,Fukuyama:2005gg}.
One remarkable feature of these models is that the scale of neutrino
masses is independent of the $(B-L)$ breaking
scale~\cite{Malinsky:2005bi}. We assume an additional singlet
superfield $\Sigma$ without direct couplings to the usual matter
multiplets. It is the out-of-equilibrium decay of this superfield
$\Sigma$ that drives leptogenesis.  Its mass can be as low as TeV,
thus avoiding conflict with reheating bounds~\cite{Kawasaki:2004qu}.
Moreover, in contrast to the simplest, unextended seesaw, one can
naturally suppress erasure of the created asymmetry due to washout
processes without conflicting with the magnitude of neutrino masses
indicated by oscillation experiments~\cite{Maltoni:2004ei}.

We consider the symmetry breaking pattern
\begin{eqnarray}
SO(10)  & \to &
SU(3)_c \times SU(2)_L \times SU(2)_R \times U(1)_{(B-L)}
\nonumber \\  & \to &
SU(3)_c \times SU(2)_L \times U(1)_{R} \times U(1)_{B-L}
\nonumber \\  & \to &
SU(3)_c \times SU(2)_L \times U(1)_{Y}
\nonumber \\  & \to &
SU(3)_c \times U(1)_{Q} .
\end{eqnarray}
We also impose a global $U(1)_G$ symmetry, under which all three
minimal \10 matter supermultiplets in the {\bf 16} are neutral and the
gauge singlet superfields $S_i, i=1,2,3$ carry non-zero $G$-charge. In
addition, we introduce two singlets $\Sigma$ and $X$ with $G=0$ and
invariant under $D-$parity, coupled to each ther so that both $X$ and
$\Sigma$ can be produced thermally and $\Sigma$ picks up a mass, from
the vacuum expectation value (vev) of the scalar in $X$.  Note that
$\Sigma$ does not couple directly with any other fields we shall just
consider in what follows $\Sigma$ as having simply a bare mass term
$M_\Sigma$.

For the symmetry breaking we consider the minimum number of Higgs
scalars. In addition to the adjoint, we break the group SO(10) with a
210-representation, which also contains a $D-$parity odd singlet
$\sigma \equiv (1,1,1,0) \subset {\bf 210}$.
The left-right symmetry is broken by a $16-$plet of Higgs (this
contains $\chi_R$ and $\chi_L$) with $G$-charge opposite to that of
the singlet matter fields $S_i$. The electroweak symmetry is broken by
a $10-$plet ($\phi$) of SO(10), neutral under $G$, which contains the
usual bi-doublet field.
Under the left-right symmetric subgroup ${\cal G}_{LR} \subset$ \10
the transformations of the remaining fields responsible for symmetry
breaking are $\phi \equiv (1,2,2,0) \subset {\bf 10}$, $\chi_R \equiv
(1,1,2,1) \subset {\bf 16}$ and $\chi_L \equiv (1,2,1,1) \subset {\bf
  16}$.  The electric charge assignment and $U(1)$ normalization are,
$$ Q = T_{3L} + T_{3R} + {B-L \over 2} = T_{3L} + {Y \over 2}.$$

The Yukawa couplings relevant for neutrino masses are
\begin{equation}
{\cal L}_Y = Y_{ij} {\nu^c}_{iL} \nu_{jL} \phi + F_{ij} \nu_{iL}
S_j \chi_L + \tilde{F}_{ij}  {\nu^c}_{iL} S_j \chi_R + M_\Sigma \Sigma \Sigma.
\end{equation}
Note that a direct Majorana mass term for the singlet fields $S_i$ is
forbidden by the $U(1)_G$ quantum number and the fact that the only
singlet scalar $\sigma$ is odd under D-parity, while $S_i S_j$ are
even under $D-parity$. For the same reason, $\sigma$ cannot couple to
$\Sigma$, although a bare mass for $\Sigma$ is allowed. This mass can
be of the order of TeV.  We also introduce a soft term breaking
$U(1)_G$, which allows mixing between these fields $\Sigma S_i$.

This will then give  a $10 \times 10$ neutrino mass matrix, in the
basis ($\nu_i$, $\Sigma$, $\nu^{c}_i$, $S_i$):
\begin{equation}
\label{eq:nu-mass-mat}
M_{\nu}=\left(
\begin{array}{cccc}
0 & 0 &Y v & F v_{L}  \\
0 & M_{\Sigma} & 0 & \Delta_{}^{T} \\
Y^{T} v & 0 & 0 & \tilde{F} v_{R} \\
F^{T} v_{L} & \Delta_{} & \tilde{F}^{T}v_{R} & 0 \\
\end{array}
\right)
\end{equation}
where, $v = \vev{\phi}$, $v_L = \vev{\chi_L}$ and $v_R = \vev{\chi_R}$
are the vevs for the fields $\phi$, $\chi_L$ and $\chi_R$ respectively
and $ \Delta_{}$ is the $U(1)_G$ breaking entry.  This mass matrix
will give two heavy states which are dominantly the right-handed
neutrino ${\nu^c}_{iL}$ and the singlets $S_i$, with a lighter state
$\Sigma$.  Using the seesaw diagonalization prescription given in
Ref.~\cite{schechter:1982cv} we obtain the effective left-handed light
neutrino mass matrix as
\begin{equation}
\label{eq:mnu}
m_{\nu} = \frac{1}{M_{\Sigma}}G G^{T}-\left[Y (F \tilde{F}^{-1})^{T}+(F \tilde{F}^{-1})Y^{T}\right]\frac{v v_{L}}{v_{R}} 
\end{equation}
where $G \equiv Y (\tilde{F}^{-1})^{T}\frac{v \Delta_{}}{v_{R}}$.

The first contribution in eq.~(\ref{eq:mnu}) arises from the soft
$U(1)_G$ breaking term.  In order to keep the absolute neutrino mass
scale in the eV range one should require
\begin{equation}
  \label{eq:bound}
\frac{v^{2}|\Delta_{}|^{2}}{M_{\Sigma}v_{R}^{2}}\lesssim {\rm eV},
\end{equation}
indicating the need for the smallness of G-violation.

We now turn to the second term. First note that its structure is
different from the conventional seesaw, first that it is linear in the
Yukawa coupling $Y$~\cite{Malinsky:2005bi}. In order to discuss its
magnitude we consider the minimization of the most general scalar
potential. This will determine the vevs of the different fields:
\[ \vev{\phi} = v; ~~~ \vev{ \chi_L} = v_L; ~~~
\vev{ \chi_R} = v_R; ~~~ \vev{ \sigma} = \eta. \] 

In models of $D-$parity violation it is usual to choose the parameters
of the potential to make the masses of the left-handed and
right-handed fields different.  A similar prescription also holds in
the presence of superymmetry, so that we can have $D-$parity violation
at a high scale, whereas the $B-L$ symmetry is broken at a scale that
can be as low as the electroweak symmetry breaking scale.
Since the $D-$parity breaking scale is much higher than the scale at
which the left-right symmetry breaks, and this in turn is higher than
the electroweak symmetry breaking scale, one has the ``vev-seesaw''
relation
\begin{equation}
\label{eq:vev-seesaw}
v_L \propto {v_R v \over M_X },
\end{equation}
where $ M_X$ is determined by the \10 breaking vevs, so that the
second contribution to the neutrino mass in Eq.~(\ref{eq:mnu}) becomes
naturally small, suppressed by the unification scale, irrespective of
the $(B-L)$ violating scale $v_R$~\cite{Malinsky:2005bi} which can be
rather low~\cite{Malinsky:2005bi}. This is in sharp contrast to the
conventional left-right symmetric seesaw models.

Note that in the present model we have a $U(1)_G$ global symmetry,
which is broken by the vev $\vev{\chi_R}$ and also explicitly through
the soft $\Sigma S$ bilinear mixing terms~\footnote{The corresponding
  would-be Goldstone picks up a large mass and/or can be made
  invisible in case all G-breaking comes spontaneously.}.

All in all, one can have naturally small neutrino masses independent
of the magnitude of the $(B-L)$ symmetry breaking scale, which may be
as low as the TeV scale.

We now discuss the issue of leptogenesis in this model.  It can occur
only after the local $(B-L) \subset$ \10 symmetry is broken. It will
take place through the decay of the singlet fermion $\Sigma$.
In order to get the total width of $\Sigma$ decaying to a lepton-Higgs
pair via the mixing with the $\nu^{c}$ and $S$ fields one should
transform the relevant superpotential term from the defining basis
($\nu$, $\Sigma$, $\nu^{c}$, $S$) to the physical matter and Higgs
doublet fields and identify the effective Yukawa coupling of $\Sigma$
to the light lepton-Higgs pair $LH$.  This way the total width of
$\Sigma$ is given by (treating $Y_{\Sigma}$ as a column vector)
\begin{equation}
  \label{eq:wid}
\Gamma_{\Sigma}\sim \frac{1}{8\pi}Y_{\Sigma}^{\dagger}Y_{\Sigma} M_{\Sigma}  
\end{equation}

In order to estimate $Y_{\Sigma}$ we need to compute the projection
$U_{\nu^{c}\Sigma}$ of $\Sigma$ onto the $\nu^c$'s. This will
determine the relevant effective coupling of the $\Sigma$ to $LH$
pair~\footnote{The ${\cal O}(1)$ coefficient $\alpha_{H}$ denotes the
  projection of the relevant light MSSM Higgs doublet $h$ into the
  directions of the defining (up-type) Higgs doublets living in $H\in
  10_{H}$. We leave it unspecified as the full-featured analysis of
  the Higgs potential is out of the scope of this work.}.

\begin{figure}[h]
\centering 
\includegraphics[height=2.6cm,width=0.5\linewidth]{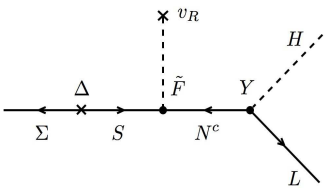}\\
\caption{Lowest order tree-level graphs giving rise to the effective
  $\Sigma L H$ Yukawa interaction; for more details see text.}
\label{fig:effectiveYSigma}
\end{figure}

In order to do that let us use again the perturbative seesaw
diagonalization prescription of Ref.~\cite{schechter:1982cv}. The
method is especially convenient to the discussion of leptogenesis, as
it includes all CP phases. One finds, at leading order:
\[
U_{\nu^{c}\Sigma}  =  (\tilde{F}^{-1})^{T}\frac{\Delta_{}}{v_{R}}, ~~~~~    
Y_{\Sigma}=\alpha_{H} Y (\tilde{F}^{-1})^{T}\frac{\Delta_{}}{v_{R}} 
\]
Note that the $G$-breaking quantity $\Delta_{}$ determines
$U_{\nu^c\Sigma}$ and $Y_{\Sigma}$.  This can be easily understood
from the Feynman diagrams for the effective Yukawa coupling
$Y_{\Sigma}$, c.f.  Fig.  \ref{fig:effectiveYSigma}.
The figure illustrates the (lowest order) tree-level graph giving rise
to the effective $\Sigma L H$ Yukawa interaction. Notice that this
decay can occur out-of-equilibrium for moderately large values of the
Yukawa couplings since, for sufficiently small values of the parameter
$\Delta_{}$, the effective $L H \nu^{c}$ vertex will be suppressed, as
seen from the graph.

The decay width of $\Sigma$ is estimated as
\begin{equation}
  \label{eq:gamma}
  \Gamma_\Sigma = {\alpha_H^2 \over 8 \pi} \frac{\Delta^\dagger}{v_{R}} (\tilde{F}^{-1})^\dagger Y^\dagger Y  (\tilde{F}^{-1}) \frac{\Delta_{}}{v_{R}} M_\Sigma  
\end{equation}

\begin{figure}[h] \centering
\includegraphics[height=2.3cm,width=.32\linewidth]{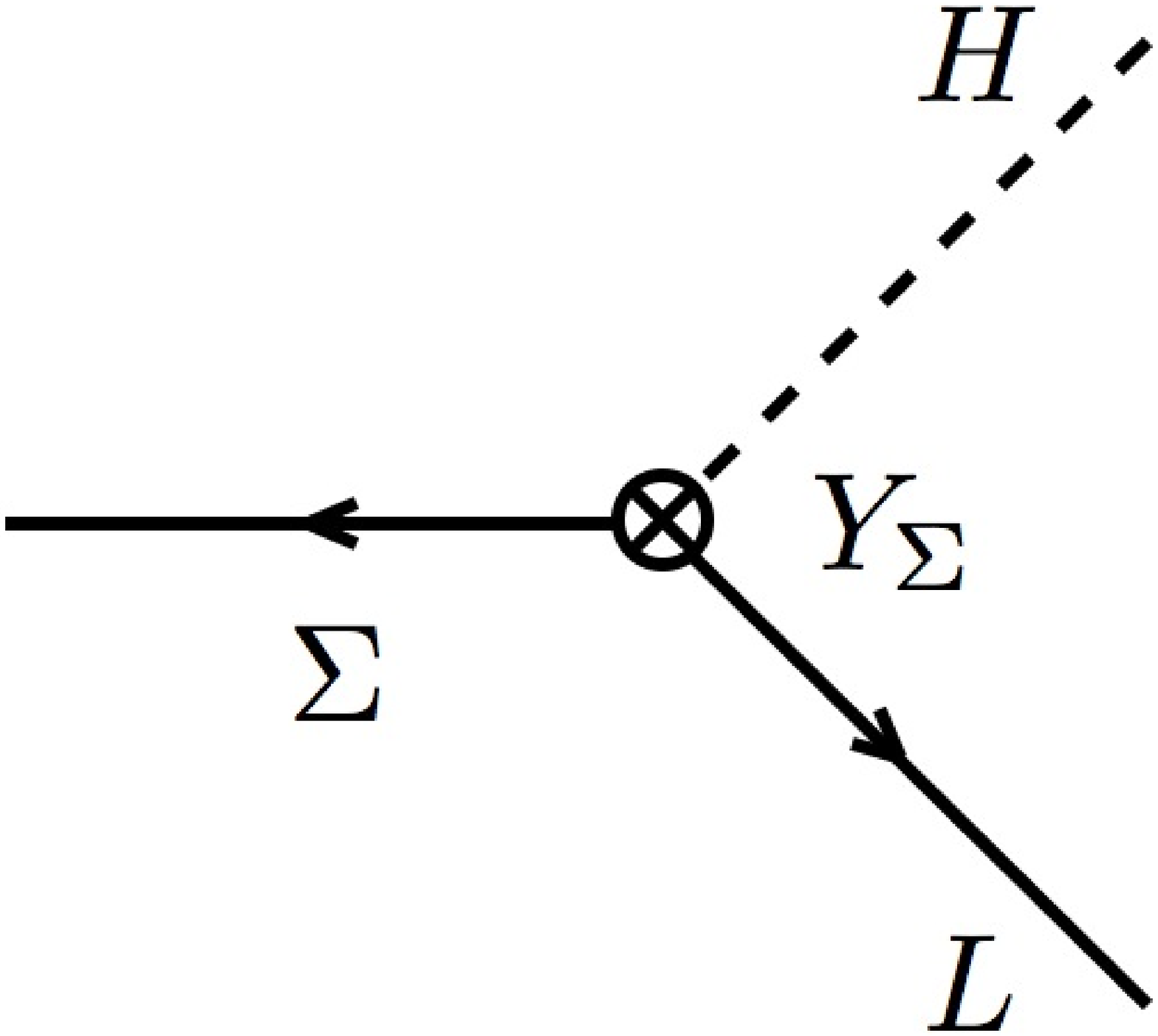}\,\,\,\,\,\,\,
\includegraphics[height=2.3cm,width=.45\linewidth]{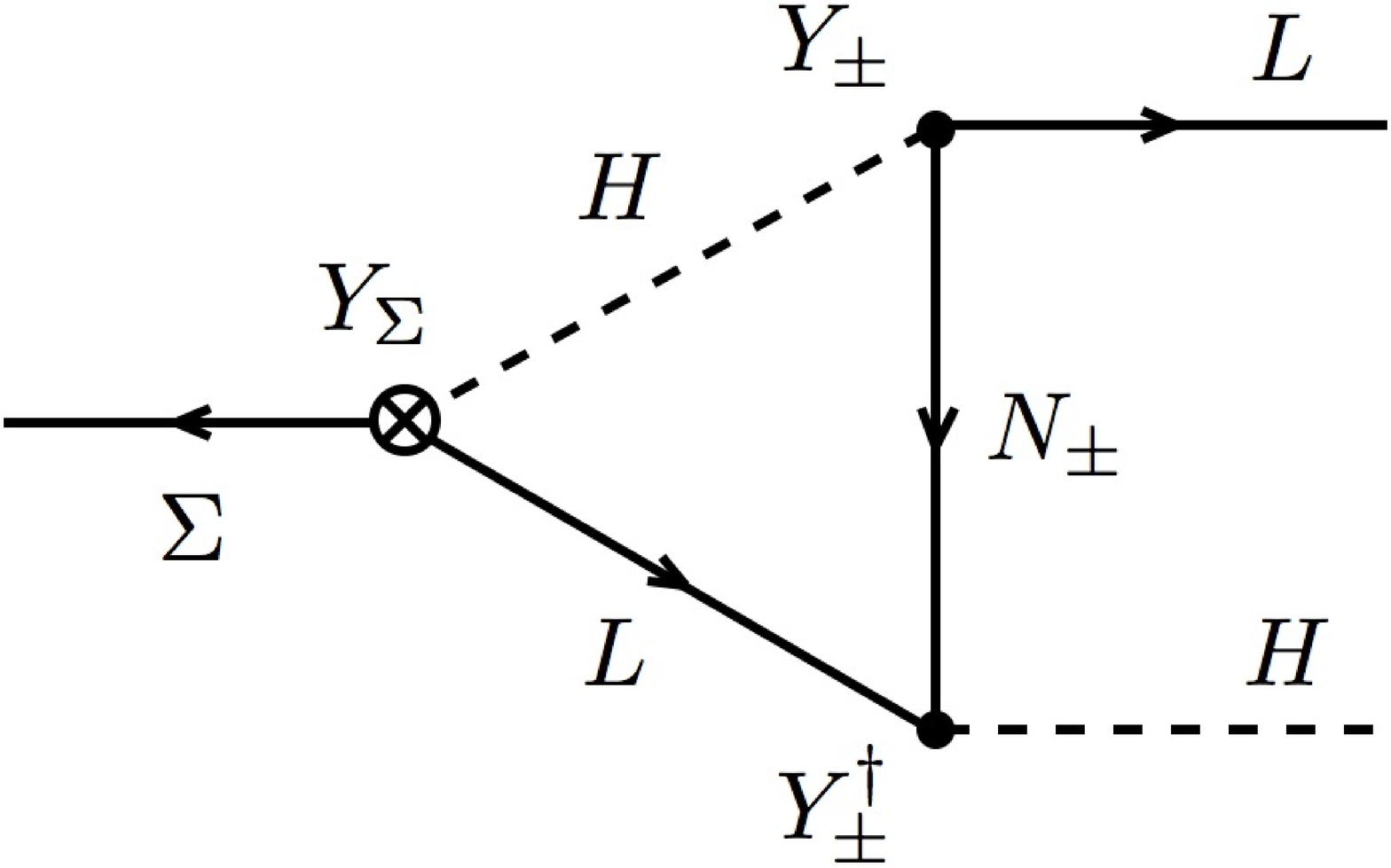}
\includegraphics[height=2.3cm,width=.51\linewidth]{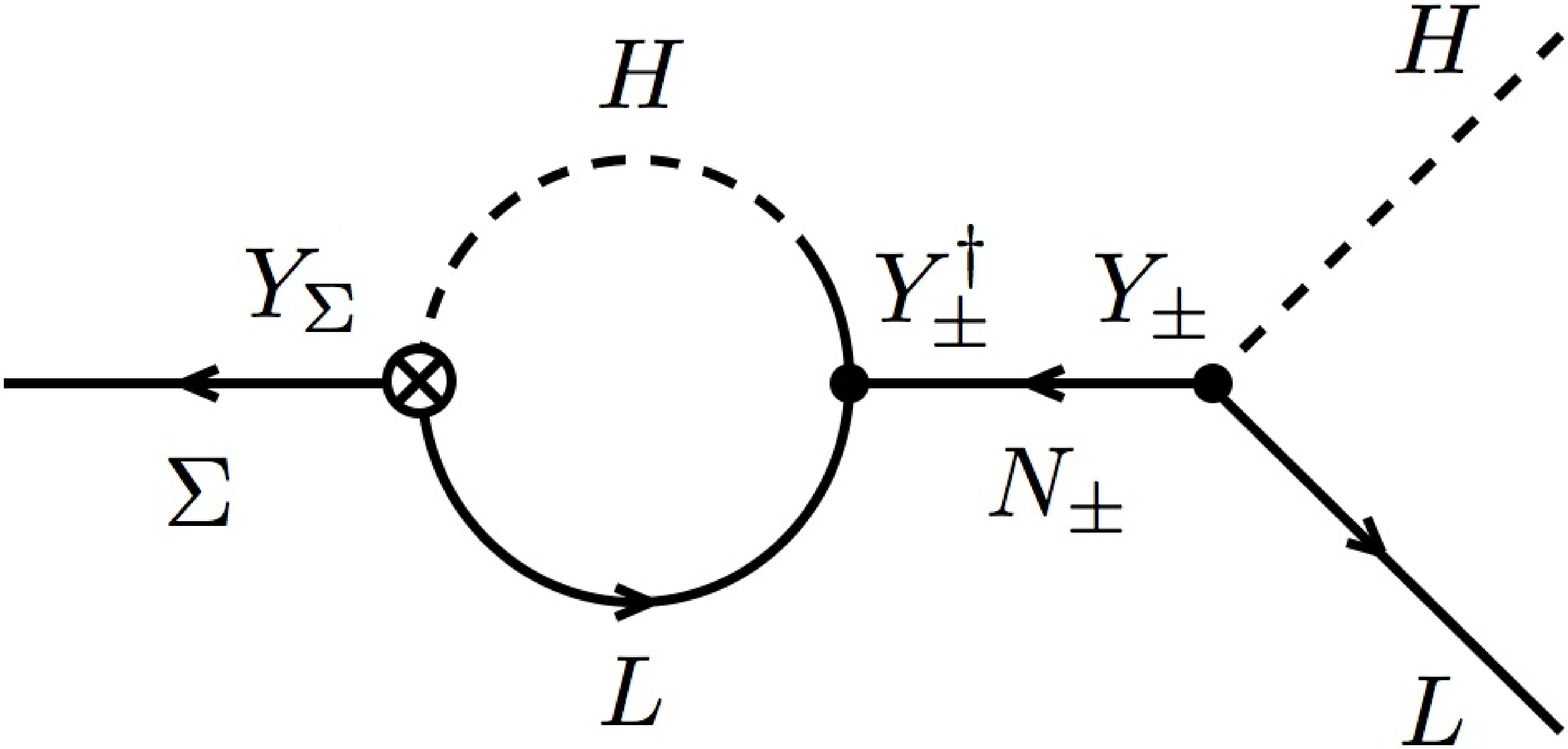}
\caption{\label{fig:l-g} Tree level and one loop diagrams for the
  decay of $\Sigma$ that interferes to generate a lepton asymmetry of
  the universe.}
\end{figure}

The interference of one loop diagrams and tree level diagrams (see
Fig.~\ref{fig:l-g}) generates a lepton asymmetry.
Keeping only the contribution of the lightest $N_1^\pm$ pair we obtain
the following estimate for the CP asymmetry produced in the decay of
$\Sigma$,
\begin{equation}
  \label{eq:asym}
\eps_{\Sigma}\propto -\frac{3}{16\pi}\frac{M_{\Sigma}}{M_{1}}\frac{{\rm Im}[(Y_{\Sigma}^{\dagger}F_{k}U_{S}\alpha_{\chi^{k}})_{1}(Y_{\Sigma}^{\dagger}Y U_{N}\alpha_{H})_{1}]}{Y^{\dagger}_{\Sigma}Y_{\Sigma}},
\end{equation}
where $\alpha_{\chi^{k}}$ are the projections of the light MSSM-like
Higgs doublet onto the defining Higgs doublets in the
$\overline{16}_{H}^{k}$ and $M_{1}$ is the mass of the (almost
degenerate) lightest pair of the
$N_{\pm}\equiv\frac{1}{\sqrt{2}}(\nu^{c}\pm S)$ states.
In addition
\begin{equation}
  \label{eq:2}
U_{N}^{T}\tilde{F}v_{R}U_{S}={\rm diag}(M_{1},M_{2},M_{3}),  
\end{equation}
The main feature of this scenario is that the suppression factor of
$\Delta/v_R$ does not enter in the amount of asymmetry generated in
the decays of $\Sigma$.
At the time of decays of $\Sigma$, the number densities of the
right-handed neutrinos and the singlets $S_i$ should be fairly less so
that they do not wash out the asymmetry generated by the decay of
$\Sigma$, which is subsequently converted to a baryon asymmetry by the
sphaleron processes.  Since the neutrino masses are maintained small
by the scale of D-parity violation, and thermal production of $\Sigma$
depends only on its coupling to $X$, there is no restriction from
neutrino masses on the couplings $Y_\Sigma$, which are dependent on
the scale of $B-L$ violation.

In order to induce successful leptogenesis the $\Sigma$ must decay
before the electroweak phase transition. Moreover $\Sigma$ decay must
take place out-of-equilibrium, i.~e. one must fulfill the condition $
\Gamma_{\mathrm{Hubble}} > \Gamma_\Sigma >
\Gamma_{\mathrm{sphaleron}}$.

\begin{figure}[h]\centering
  \includegraphics[height=5cm,width=.85\linewidth]{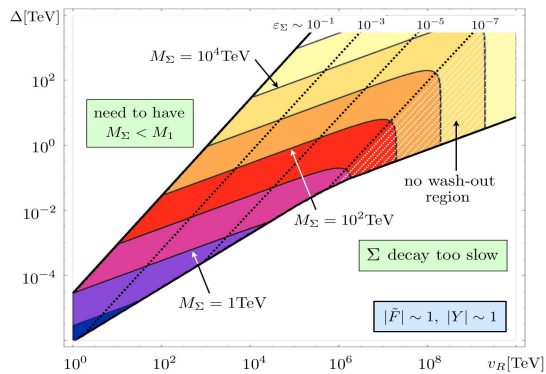}
  \caption{Leptogenesis parameter region (see text).}
 \label{fig:correlation1}
\end{figure}

Fig.~\ref{fig:correlation1} shows the typical correlations among the
magnitudes of parameters $v_{R}$, $\Delta$ and $M_{\Sigma}$ leading to
roughly $\eta \sim \eps_{\Sigma} \kappa/g_{*} \sim 6 \times 10^{-10}$, where $g_{*}$ is the relevant number of degrees of freedom $\sim 2\times 10^{2}$,
$\kappa$ is 1 if the width is well below the Hubble rate, and falls
exponentially otherwise~\footnote{The lepton asymmetry $Y_L = \epsilon
  \kappa/g_*$ is defined as $(n_L-n_{\bar L})/s$ with $s$ being the
  entropy.  The
  conversion of L into B brings another factor 1/3 or so.  }. In our
estimates we assume all Yukawas are order unity, e.g.
$|\tilde{F}|\sim |Y|\sim 1$.
For a given $v_{R}$ there is only a certain range for $M_{\Sigma}$: i)
the lower bound indicated by the lower solid line comes from the need
to generate enough asymmetry (proportional to
$M_{\Sigma}/|\tilde{F}|v_{R}$) while ii) the upper bound stems from
the need to have $M_{\Sigma}$ below $M_{1}$. 
Note that in the upper right region wash-out is negligible, here the
asymmetry $\eps_{\Sigma}$ is essentially constant as $\sim
M_{\Sigma}/|\tilde{F}|v_{R}$. On the other hand, in the region left of
the dotted line with $\eps_{\Sigma}=10^{-7}$ one would have too large
an asymmetry, $\eps_{\Sigma}>10^{-7}$, were it not for the fact that,
in this region, this is compensated by a certain amount of wash-out,
so as to lead to an acceptable asymmetry.
Note also the relative smallness of the G-breaking $\Delta$ parameter.
One sees, for example, that successful leptogenesis can occur for
$M_\Sigma = 1$ TeV and low $v_R = 10$ TeV.

In short, we have considered a supersymmetric \10 seesaw model with
only doublet Higgs scalars, in which neutrino masses are suppressed by
the scale of D-parity violation, regardless of the value of the
$(B-L)$ violating scale, which can be low. This would allow for the
existence of new physics (e.~g. a $Z^\prime$ gauge boson) accessible
at accelerators. Leptogenesis can occur at the TeV scale through the
decay of a singlet $\Sigma$, thereby avoiding the gravitino crisis.
Washout of the asymmetry is suppressed by the absence of direct
couplings of $\Sigma$ to leptons.

Note that the mechanism described here involving the addition of the
$\Sigma$ field is very natural in the framework of the extended seesaw
model but not in the simplest type-I seesaw scheme~\cite{Ma:2006ci}.
Details of the mechanism and a critical comparison with unextended
seesaw schemes will be presented elsewhere~\cite{preparation}.

\def\baselinestretch{1}%
Work supported by MEC grants FPA2005-01269 and BFM2002-00345, by EC
Contracts RTN network MRTN-CT-2004-503369 and ILIAS/N6
RII3-CT-2004-506222.  M.~H.  is supported by a Ramon y Cajal grant.

 \def\baselinestretch{1}%


\begin{thebibliography}{10}

\bibitem{Fukugita:1986hr}
M.~Fukugita and T.~Yanagida,
\newblock Phys. Lett. {\bf B174}, 45 (1986).

\bibitem{Buchmuller:2005eh}
W.~Buchmuller, R.~D. Peccei and T.~Yanagida,
\newblock Ann. Rev. Nucl. Part. Sci. {\bf 55}, 311 (2005), [hep-ph/0502169].

\bibitem{Minkowski:1977sc}
P.~Minkowski,
\newblock Phys. Lett. {\bf B67}, 421 (1977).

\bibitem{Orloff:2005nu} Articles by M. Gell-Mann, P. Ramond and R.
  Slansky; T.  Yanagida; R. Mohapatra and G. Senjanovic in Proc. of
  Int. Conf. on the Seesaw Mechanism and the Neutrino Mass, Paris,
  France, 10-11 June 2004, edited by J.~Orloff, S.~Lavignac and
  M.~Cribier.

\bibitem{schechter:1980gr}
J.~Schechter and J.~W.~F. Valle,
\newblock Phys. Rev. {\bf D22}, 2227 (1980).

\bibitem{schechter:1982cv}
J.~Schechter and J.~W.~F. Valle,
\newblock Phys. Rev. {\bf D25}, 774 (1982).

\bibitem{Lazarides:1980nt}
G.~Lazarides, Q.~Shafi and C.~Wetterich,
\newblock Nucl. Phys. {\bf B181}, 287 (1981).

\bibitem{Buchmuller:2004nz}
W.~Buchmuller, P.~Di~Bari and M.~Plumacher,
\newblock  Annals Phys.\  {\bf 315} (2005) 305 [hep-ph/0401240].

\bibitem{Giudice:2003jh}
G.~F. Giudice et al,
\newblock Nucl.\ Phys.\ B {\bf 685} (2004) 89 

\bibitem{Khlopov:1984pf}
M.~Y. Khlopov and A.~D. Linde,
\newblock Phys. Lett. {\bf B138}, 265 (1984).

\bibitem{Kawasaki:2004qu}
M.~Kawasaki, K.~Kohri and T.~Moroi,
\newblock Phys. Rev. {\bf D71}, 083502 (2005), [astro-ph/0408426].

\bibitem{Farzan:2005ez}
Y.~Farzan and J.~W.~F. Valle,
\newblock Phys. Rev. Lett. {\bf 96}, 011601 (2006), [hep-ph/0509280].

\bibitem{Malinsky:2005bi}
M.~Malinsky, J.~C. Romao and J.~W.~F. Valle,
\newblock Phys. Rev. Lett. {\bf 95}, 161801 (2005), [hep-ph/0506296].

\bibitem{mohapatra:1986bd}
R.~N. Mohapatra and J.~W.~F. Valle,
\newblock Phys. Rev. {\bf D34}, 1642 (1986).

\bibitem{Akhmedov:1995vm}
E.~Akhmedov, M.~Lindner, E.~Schnapka and J.~W.~F. Valle,
\newblock Phys. Rev. {\bf D53}, 2752 (1996), [hep-ph/9509255].

\bibitem{Barr:2005ss}
S.~M. Barr and I.~Dorsner,
\newblock Phys. Lett. {\bf B632}, 527 (2006) 

\bibitem{Fukuyama:2005gg}
T.~Fukuyama, A.~Ilakovac, T.~Kikuchi and K.~Matsuda,
\newblock JHEP {\bf 06}, 016 (2005), [hep-ph/0503114].

\bibitem{Maltoni:2004ei}
M.~Maltoni, T.~Schwetz, M.~A. Tortola and J.~W.~F. Valle,
\newblock New J. Phys. {\bf 6}, 122 (2004),
\newblock Appendix C in hep-ph/0405172 (v5) provides updated results which take
  into account all developments as of June 2006, namely: new SSM, new SNO salt
  data, latest K2K and MINOS data; previous works by other groups are
  referenced therein.

\bibitem{Ma:2006ci}
  E.~Ma, N.~Sahu and U.~Sarkar,
  arXiv:hep-ph/0603043.

\bibitem{preparation}
M.~Malinsky et al, in preparation.

\end{thebibliography}
\end{document}